\documentclass{revtex4}

\usepackage{graphicx}
\begin{document}
% You should use BibTeX and revtex.bst for references
\bibliographystyle{apsrev}

% Use the \preprint command to place your local institutional report
% number  and your conference paper identification number on the
% title page in preprint mode. Multiple \preprint commands are allowed.
\preprint{BUHEP-01-23}
\preprint{HUTP-01/A043}

%Title of paper
\title{Technicolor Evolution}
% Optional argument for running titles on pages
%\title[]{Technicolor Evolution}

% repeat the \author .. \affiliation  etc. as needed
% \email, \thanks, \homepage, \altaffiliation all apply to the current
% author. Explanatory text should go in the []'s, actual e-mail
% address or url should go in the {}'s for \email and \homepage.
% Please use the appropriate macro for the type of information

% \affiliation command applies to all authors since the last
% \affiliation command. The \affiliation command should follow the
% other information

\author{Elizabeth H. Simmons}
\email[]{simmons@bu.edu}
\homepage[]{http://smyrd.bu.edu/mypage/mypage.html}
\thanks{Talk presented at the Summer Study on the Future of Particle Physics, Snowmass, CO, June 30 - July 21, 2001. Preprint numbers: BUHEP-01-23, HUTP-01/A043.}
%\altaffiliation{}
\affiliation{Physics Department, Boston University, 590 Commonwealth Avenue, Boston, MA   02215\\
Radcliffe Institute for Advanced Study and Department of
Physics, Harvard University, Cambridge,  MA   02138}

%Collaboration name if desired (requires use of superscriptaddress
%option in \documentclass). \noaffiliation is required (may also be
%used with the \author command).
%\collaboration{}
%\noaffiliation

\date{\today}

\begin{abstract} 
This talk describes how modern theories of dynamical electroweak
symmetry breaking have evolved from the original minimal QCD-like
technicolor model in response to three key challenges: $R_b$,
flavor-changing neutral currents, and weak isospin violation.
\end{abstract}
% insert suggested PACS numbers in braces on next line
% \pacs{12.60.Nz,12.60.Cn,12.60.-i}

%\maketitle must follow title, authors, abstract and \pacs
\maketitle

% body of paper here - Use proper section commands
% References should be done using the \cite, \ref, and \label commands
%\section{}
%\label{}
%\subsection{}
%\subsubsection{}

\section{Introduction}

\begin{figure}[bt]
\begin{center}
\scalebox{.36}{\includegraphics{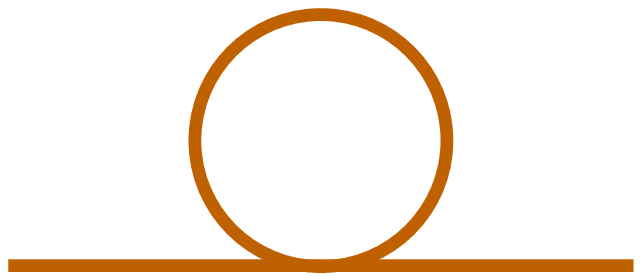}}
\hspace{3cm} \scalebox{.4}{\includegraphics{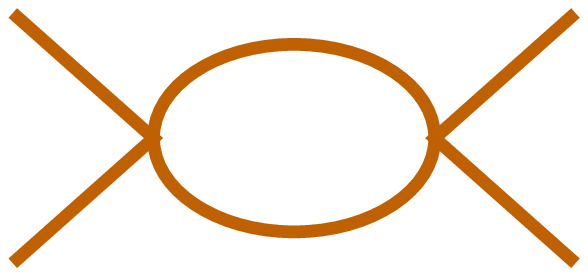}}
\end{center}
\caption{(left) Naturalness problem: $ M_H^2\ \propto\ \Lambda^2$. (right)
Triviality: $\beta(\lambda)\ = \ {{3\lambda^2}\over{2\pi^2}}\ > \
  0$.}
\label{ehs:fig1}
\end{figure}

In order to understand the origin of mass, we must find both the cause
of electroweak symmetry breaking, through which the $W$ and $Z$ bosons
obtain mass, and the cause of flavor symmetry breaking, by which the
quarks and leptons obtain their diverse masses and mixings.  The
Standard Higgs Model of particle physics, based on the gauge group
$SU(3)_c \times SU(2)_W \times U(1)_Y$ accommodates both symmetry
breakings by including a fundamental weak doublet of scalar
(``Higgs'') bosons ${\phi = {\phi^+ \choose \phi^0}}$ with potential
function $V(\phi) = \lambda \left({\phi^\dagger \phi - \frac12
v^2}\right)^2$.  However the Standard Model does not explain the
dynamics responsible for the generation of mass.  Furthermore, the
scalar sector suffers from two serious problems.  The scalar mass is
unnaturally sensitive to the presence of physics at any higher scale
$\Lambda$ (e.g. the Planck scale), as shown in fig. \ref{ehs:fig1}.
This is known as the gauge hierarchy problem.  In addition, if the
scalar must provide a good description of physics up to arbitrarily
high scale (i.e., be fundamental), the scalar's self-coupling
($\lambda$) is driven to zero at finite energy scales as indicated in
fig. \ref{ehs:fig1}.  That is, the scalar field theory is free (or
``trivial''). Then the scalar cannot fill its intended role: if
$\lambda = 0$, the electroweak symmetry is not spontaneously broken.
The scalars involved in electroweak symmetry breaking must therefore
be a party to new physics at some finite energy scale -- e.g., they
may be composite or may be part of a larger theory with a UV fixed
point.  The Standard Model is merely a low-energy effective field
theory, and the dynamics responsible for generating mass must lie in
physics outside the Standard Model.

\noindent 

This talk focuses on Dynamical Electroweak Symmetry Breaking, an
approach that evades the hierarchy and triviality problems because the
scalar states involved in electroweak symmetry breaking are manifestly
composite at scales not much above the electroweak scale $v \sim 250$
GeV. .  The prototypical model of this kind is Technicolor, originally
conceived as having dynamics modeled on those of QCD.  To build a
minimal technicolor model, one starts with the Standard Model, removes
the Higgs doublet, and adds an asymptotically free (technicolor) gauge
force, e.g. based on the group $SU(N)_{TC}$, and a set of massless
technifermions which feel this new force.  The electroweak charges of
the technifermions are chosen so that the formation of a technifermion
condensate will break the electroweak symmetry to its electromagnetic
subgroup.  The simplest choice is to include two flavors of
technifermions, of which the right-handed components, $U_R$ and $D_R$,
are weak singlets while the left-handed members form a weak doublet
$(U, D)_L$.  The Lagrangian for the technifermions, like that of
massless up and down quarks, possesses a global chiral symmetry
$SU(2)_L \times SU(2)_R$.  At a scale $\Lambda_{TC} \sim 1$ TeV, the
technicolor coupling $g_{TC}$ becomes strong, causing the
technifermions to condense: $\langle \bar{U}U + \bar{D}D\rangle \neq
0$.  The condensate breaks the technifermions' chiral symmetries to
the vector subgroup $SU(2)_{L+R}$; the Nambu-Goldstone bosons of this
symmetry breaking are called technipions $\Pi_T$, in analogy with the
pions of QCD.  Because of the technifermions' electroweak quantum
numbers, the condensate also breaks $SU(2)_W \times U(1)_Y$ to
$U(1)_{EM}$, and the technipions become the longitudinal
modes of the $W$ and $Z$.  The logarithmic running of the strong
technicolor gauge coupling renders the low value of the electroweak
scale (i.e.  the gauge hierarchy) natural in these theories, while the
absence of fundamental scalars obviates concerns about triviality.

Even a minimal technicolor sector, as described above, should yield
visible signatures in collider experiments.  For example, the
technihadron spectrum, like the QCD hadron spectrum, will include
vector resonances like techni-rho and techni-omega states ($\rho_T,\
\omega_T$).  These should contribute to the rescattering of
longitudinal electroweak bosons radiated from initial-state quarks at
the LHC.  For relatively light states $M_{\rho_T} \sim 1 TeV$, a peak
might be visible in the invariant mass spectrum for production of
$W+Z$ where both bosons decay leptonically \cite{Bagger:1994zf,
Bagger:1995mk}.  An even sharper signal due to mixing of the $\rho_T$
with electroweak bosons could also be present in direct $q\bar{q} \to
W_L Z_L$ processes \cite{Golden:1995xv}.  While one might also have
hoped to detect technicolor through enhanced non-resonant $W^+ W^+$
scattering, the signal is neither large nor kinematically distinct
from the background \cite{Armstrong:1994it}.  At a linear $e+ e^-$
collider with $\sqrt{s} = 1.5$ TeV and ${\cal L}$ = 200 fb$^{-1}$, a 1
TeV vector resonance could make its presence felt in $e^+ e^- \to W^+
W^- \bar{\nu}\nu$ (but not in $e^+ e^- \to Z Z e^+ e^-$), while the
non-resonant ``low-energy theorem'' contribution would, again, be
undetectable \cite{Barger:1995cn}.  The $Z$-boson form factor would
also be sensitive to a $\rho_T$ with a mass up to a few TeV
\cite{Barklow:1994uf}.  Finally, at a 4 TeV muon collider with ${\cal
L}$ = 200 fb$^{-1}$, the gauge-boson rescattering process $\mu^+ \mu_-
\to W^+ W^- X$ ($Z Z X$) with hadronically-decaying $W$'s ($Z$'s) will
be sensitive to a technirho of up to 2 TeV (1 TeV); same-sign W
production from like-sign muon beams shows only a featureless increase
in the WW invariant mass \cite{Barger:1997kp}.

In order to generate masses and mixings for the quarks and leptons, it
is necessary to couple them to the source of electroweak symmetry
breaking.  The classic way of doing this is by extending the
technicolor gauge group to a larger extended technicolor (ETC) group
under which the ordinary fermions are also charged.  When ETC breaks
to its technicolor subgroup at a scale $M > \Lambda_{TC}$, the gauge
bosons coupling ordinary fermions to technifermions acquire a mass of
order $M$.  At the scale $\Lambda_{TC}$, a technifermion condensate
breaks the electroweak symmetry as described earlier, and the quarks
and leptons acquire mass because the massive ETC gauge bosons couple
them to the condensate.  The top quark's mass, e.g., arises when the
condensing technifermions transform the scattering diagram in fig. 
\ref{ehs:fig2:etccond} (left) into the top self-energy diagram shown at
right.  Its size is
\begin{equation}
m_t \approx (g_{ETC}^2/M^2)\langle T\bar T\rangle \approx 
(g_{ETC}^2/M^2)(4\pi v^3)\ \ \ .
\label{ehs:eq:mttopp}
\end{equation}
Thus $M$ must satisfy $M/g_{ETC} \approx 1.4$ TeV in order to produce
$m_t = 175$ GeV.

{\begin{figure}[t]
\begin{center}
\scalebox{.2}{\includegraphics{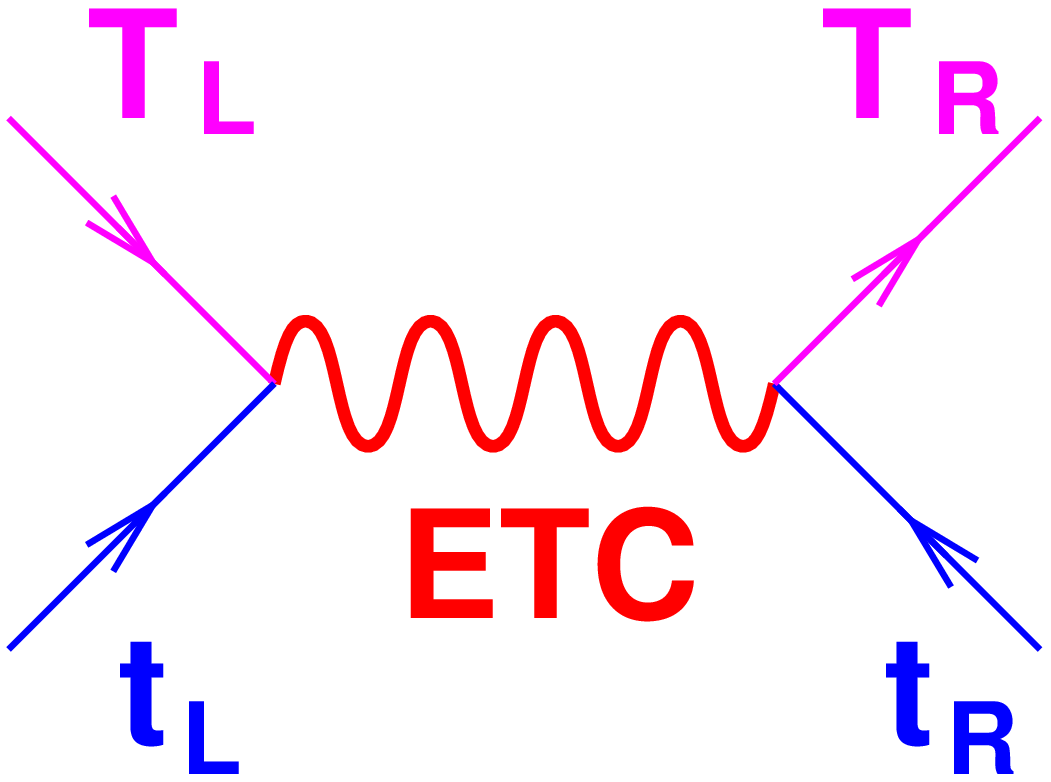}}
\hspace{2cm} \scalebox{.2}{\includegraphics{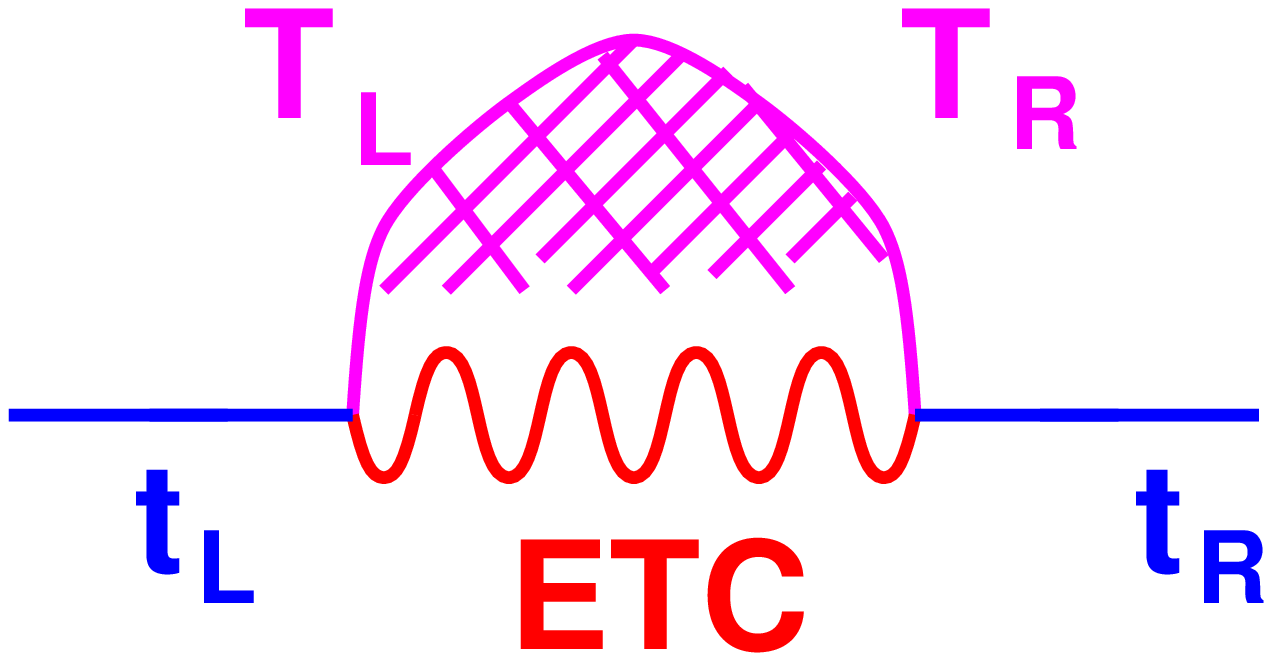}}
\end{center}
\caption{(left) Top-technifermion scattering mediated by a heavy ETC
gauge boson. (right) Technifermion condensation creates the top quark
mass.}
\label{ehs:fig2:etccond}
\end{figure}}

While this mechanism works well in principle, it has proven difficult
to construct a complete model that can accommodate the wide range of
observed fermion masses while remaining consistent with precision
electroweak data.  Three key challenges have led model-building in new
and promising directions.  First, the dynamics responsible for the
large value of $m_t$ must couple to $b_L$ because $t$ and $b$ are weak
partners.  How, then, can one obtain a predicted value of $R_b$ that
agrees with experiment?  Attempts to answer this question have led to
models in which the weak interactions of the top
quark\cite{Chivukula:1994mn, Chivukula:1996gu, Muller:1996dj,
Malkawi:1996fs} (and, perhaps, all third generation fermions) are
non-standard.  Second, while creating large fermion masses $m_f$
requires $M_{ETC}$ to be of order one TeV, suppressing flavor-changing
neutral currents demands that $M_{ETC}$ be several orders of magnitude
higher.  Attempts to resolve this conflict have led to the idea that
the technicolor gauge dynamics may have a small beta-function:
$\beta_{TC} \approx 0$ \cite{Yamawaki:1986zg,Appelquist:1986an}.  Such
``walking technicolor'' models often include light technihadrons with
distinctive signatures \cite{Eichten:1997yq}.  Third, despite the
large mass splitting $m_t \gg m_b$, the value of the rho parameter is
very near unity.  How can dynamical models accommodate large weak
isospin violation in the $t-b$ sector without producing a large shift
in $M_W$?  This issue has sparked theories in which the strong (color)
interactions of the top quark\cite{Hill:1995hp} (and possibly other
quarks\cite{Popovic:1998vb}) are modified from the predictions of QCD.

In the remainder of this talk, we explore the ways in which modern
theories of dynamical electroweak symmetry breaking have evolved in
response to these issues in flavor physics.

\section{The $R_b$ Challenge}

In classic extended technicolor models, the large value of $m_t$ comes from ETC dynamics at
a relatively low scale $M_{ETC}$ of order a few TeV.  At that scale, the weak
symmetry is still intact so that $t_L$ and $b_L$ function as weak partners.
Moreover, experiment tells us that $\vert V_{tb} \vert \approx 1$.  As a
result, the ETC dynamics responsible for generating $m_t$ must couple with
equal strength to $t_L$ and $b_L$.  While many properties of the top quark
are only loosely constrained by experiment, the $b$ quark has been far more
closely studied.  In particular, the LEP measurements of the $Zb\bar{b}$
coupling are precise enough to be sensitive to the quantum corrections
arising from physics beyond the Standard Model.  As we now discuss, radiative
corrections to the $Zb\bar{b}$ vertex from low-scale ETC dynamics can be so
large that new weak interactions for the top quark are required to make the
models consistent with experiment.\cite{Chivukula:1992ap,Chivukula:1993tz,Chivukula:1994mn}

\begin{figure}[tb]
\begin{center}
\scalebox{.2}{\includegraphics{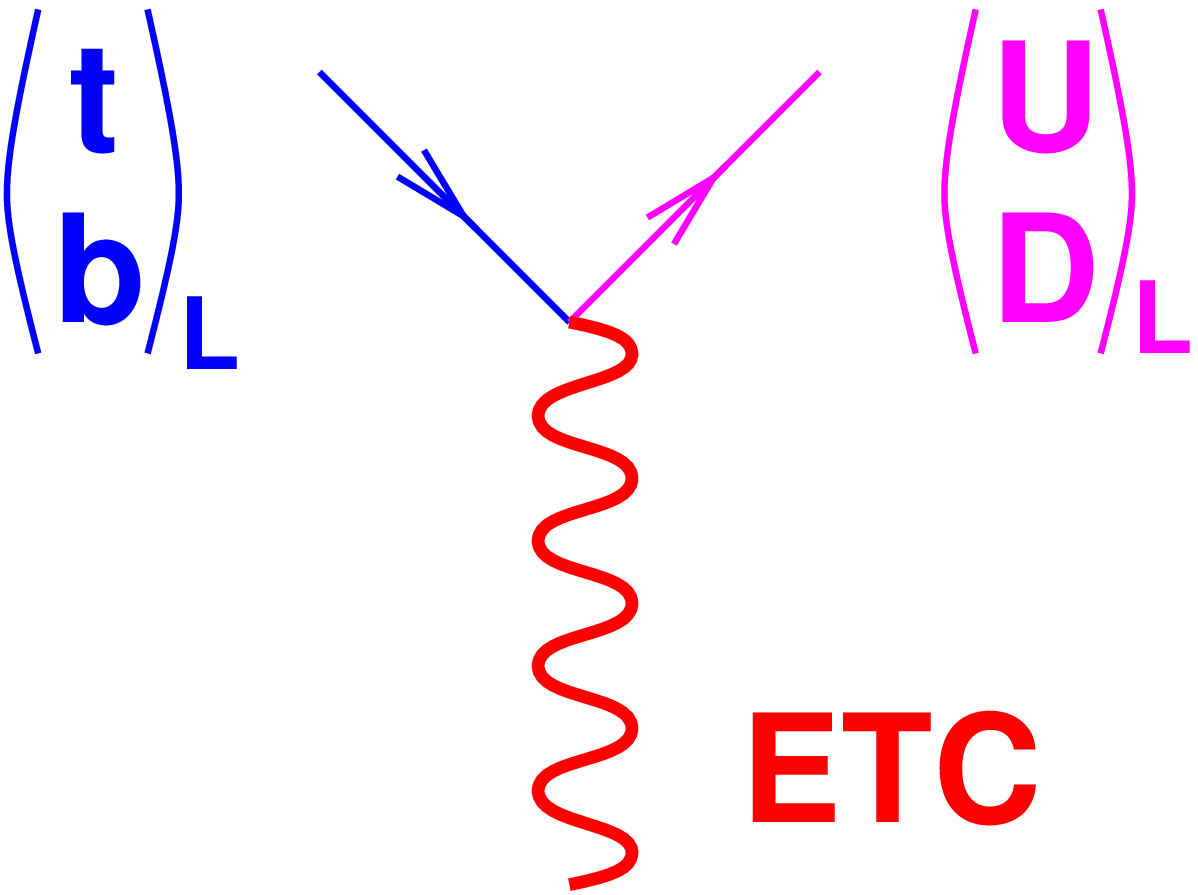}}
\hspace{2cm} \scalebox{.2}{\includegraphics{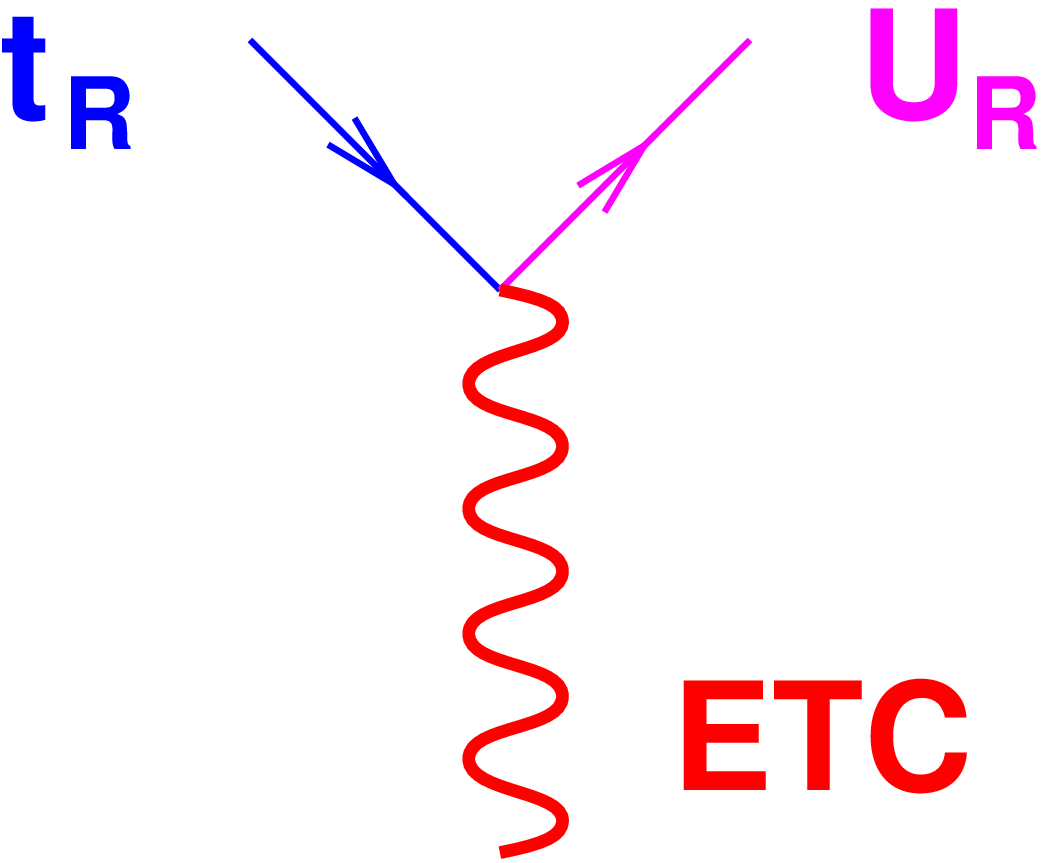}}
\end{center}
\caption{Fermion currents coupling to the weak-singlet ETC boson that
generates $m_t$.}
\label{ehs:fig3:etcfer}
\end{figure}

To begin, consider the usual ETC models in
which the extended technicolor and weak gauge groups commute, so that the
ETC gauge bosons carry no weak charge.  In these models, the ETC gauge
boson whose exchange gives rise to $m_t$ couples to the fermion
currents\cite{Chivukula:1992ap,Chivukula:1993tz} 
\begin{equation}
\xi \left(\bar\psi^i_L\ \gamma^\mu\ T^{ik}_L\right)\  +\ 
\xi^{-1} \left(\bar t_R\ \gamma^\mu\ U^k_R\right)
\end{equation}
where $\xi$ is a Clebsh of order 1 (see fig. \ref{ehs:fig3:etcfer}).
Then the top quark mass arises from technifermion condensation and ETC
boson exchange as in fig. \ref{ehs:fig2:etccond}, with the relevant
technifermions being $U_L$ and $U_R$.

\begin{figure}[tb]
\begin{center}
\scalebox{.2}{\includegraphics{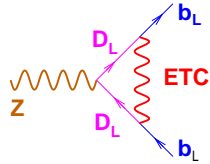}}
\end{center}
\caption[gr]{Direct correction to the $Zb\bar{b}$ vertex from the ETC gauge
boson responsible for $m_t$ in a commuting model.}
\label{ehs:fig4:etc-vv}
\end{figure}

Exchange of the same\cite{Chivukula:1992ap,Chivukula:1993tz} ETC boson causes a
direct (vertex) correction to the $Z\to b\bar{b}$ decay as shown in fig. 
\ref{ehs:fig4:etc-vv}; note that it is $D_L$ technifermions with 
$I_3 = -\frac{1}{2}$ which enter the loop.
This effect reduces the magnitude of the $Zb\bar{b}$ coupling by an amount governed by the size of the top quark mass
\begin{equation}
\delta g_L = { e \over {4 \sin{\theta} \cos{\theta}}} 
\left({g^2 v^2 \over {M^2}}\right) \approx { e \over {4 \sin{\theta} \cos{\theta}}} {m_t \over {4 {\pi} v}}
\label{ehs:eq:delgg}
\end{equation}
where we have used the relationship between $M_{ETC}$ and
$m_t$ from eqn. \ref{ehs:eq:mttopp}.

The shift in the coupling directly affects the ratio of $Z$ decay widths 
$R_b \equiv \Gamma(Z \to b\bar{b}) / \Gamma(Z \to {\rm hadrons})$, such that the fractional change in $R_b$ is proportional to $\delta g_L$.  Moreover, oblique and QCD corrections to the decay widths cancel in the ratio, up to
factors suppressed by small quark masses.  One finds\cite{Chivukula:1992ap,Chivukula:1993tz}
\begin{equation}
{{\delta R_b \over R_b} \approx - 5.1\%\cdot \xi^2\cdot  
\left(\frac{m_t}{175{\rm GeV}}\right) }
\label{ehs:eq:delrbb}
\end{equation}
Such a large shift in $R_b$ is excluded\cite{ehs_PDBook} by the data.  Then
the ETC models whose dynamics produces this shift are likewise
excluded.

This suggests one should consider an alternative class of ETC
models\cite{Chivukula:1994mn} in which the weak group $SU(2)_W$ is embedded in
$G_{ETC}$, so that the weak bosons carry weak charge.  Embedding the weak
interactions of all quarks in a low-scale ETC group would produce masses of
order $m_t$ for all up-type quarks.  Instead, one can extend $SU(2)$ to a
direct product group $SU(2)_h \times SU(2)_\ell$ such that the third
generation fermions transform under $SU(2)_h$ and the others under
$SU(2)_\ell$.  Only $SU(2)_h$ is embedded in the low-scale ETC group; the
masses of the light fermions will come from physics at higher scales.
Breaking the two weak groups to their diagonal subgroup ensures approximate
Cabibbo universality at low energies.  The electroweak and technicolor
gauge structure of these non-commuting models is sketched below\cite{Chivukula:1994mn}:
\begin{eqnarray}
G_{ETC}  &\times& SU(2)_{light} \times U(1) \nonumber\\
&\downarrow& f \nonumber \\
G_{TC} &\times& SU(2)_{heavy}  \times SU(2)_{light}  \times
U(1)_Y \nonumber\\
&\downarrow& u \nonumber\\
G_{TC}  &\times& SU(2)_{weak} \times U(1)_Y\nonumber\\
&\downarrow& v\nonumber\\
G_{TC} & \times&  U(1)_{EM}
\end{eqnarray}

Due to the extended gauge structure, three sets of electroweak gauge bosons are present in the spectrum:  heavy
states $W^H, Z^H$ that couple mainly to the third generation, light
states $W^L, Z^L$ resembling the standard $W$ and $Z$, and a massless
photon $A^\mu = \sin\theta [\sin\phi \,W_{3\ell}^\mu +
\cos\phi\,W_{3h}^\mu] +\cos\theta X^\mu $ coupling to $Q = T_{3h} +
T_{3\ell} + Y $.  Here, $\phi$ describes the mixing between the two
weak groups and $\theta$ is the usual weak angle.  The coupling of $Z_L$ to quarks differs from the standard model value by  $\delta g_L = (c^4/x) T_{3\ell} -
(c^2s^2/x) T_{3h}$.  This reduces $R_b$\cite{Chivukula:1994mn}
\begin{equation}
\frac{\delta R_b}{R_b} \approx - 5.1\% \ [\sin^2\phi \frac{f^2}{u^2}]\ 
\end{equation}
where the term in square brackets is ${\cal O}(1)$.

\begin{figure}[tb]
\begin{center}
\scalebox{.2}{\includegraphics{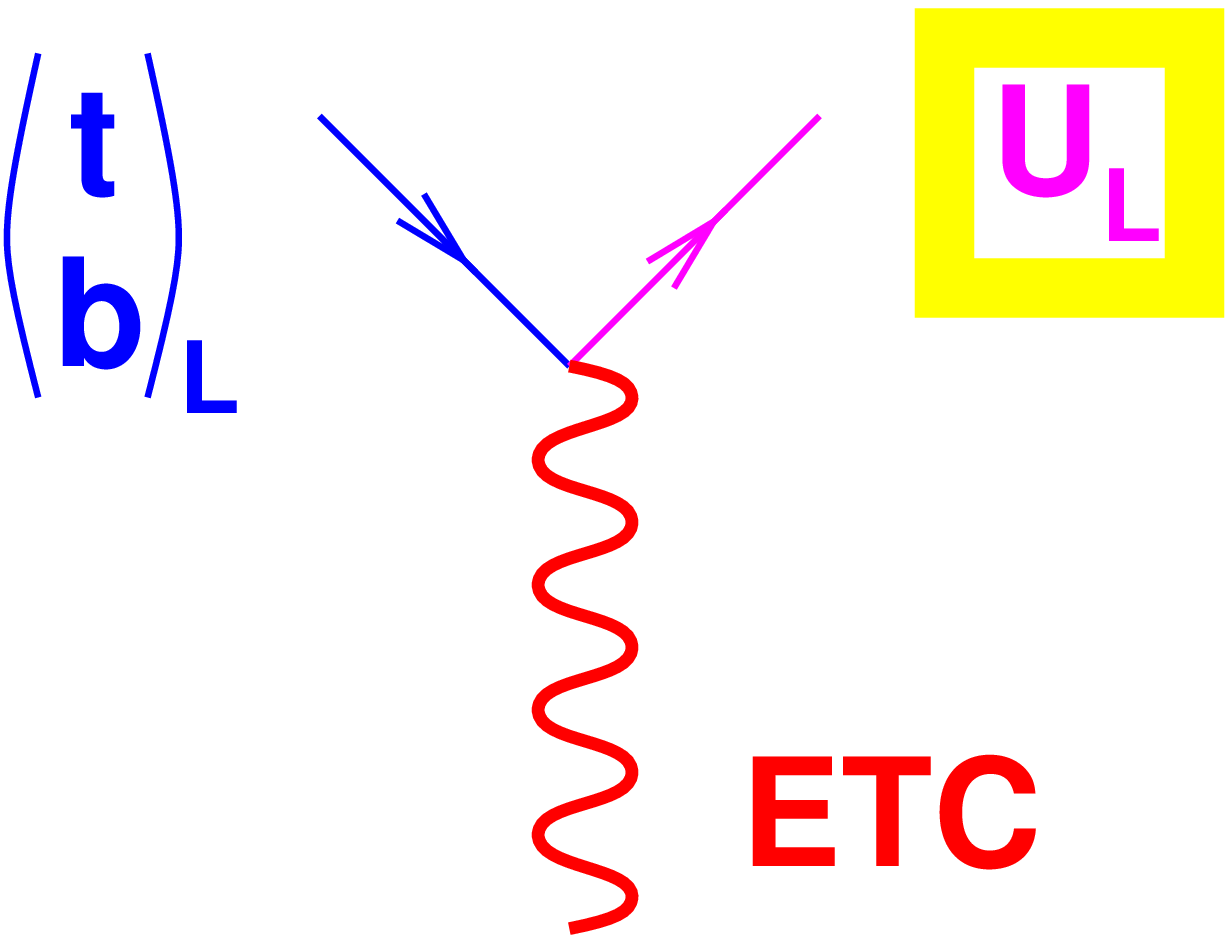}}
\hspace{2cm} \scalebox{.2}{\includegraphics{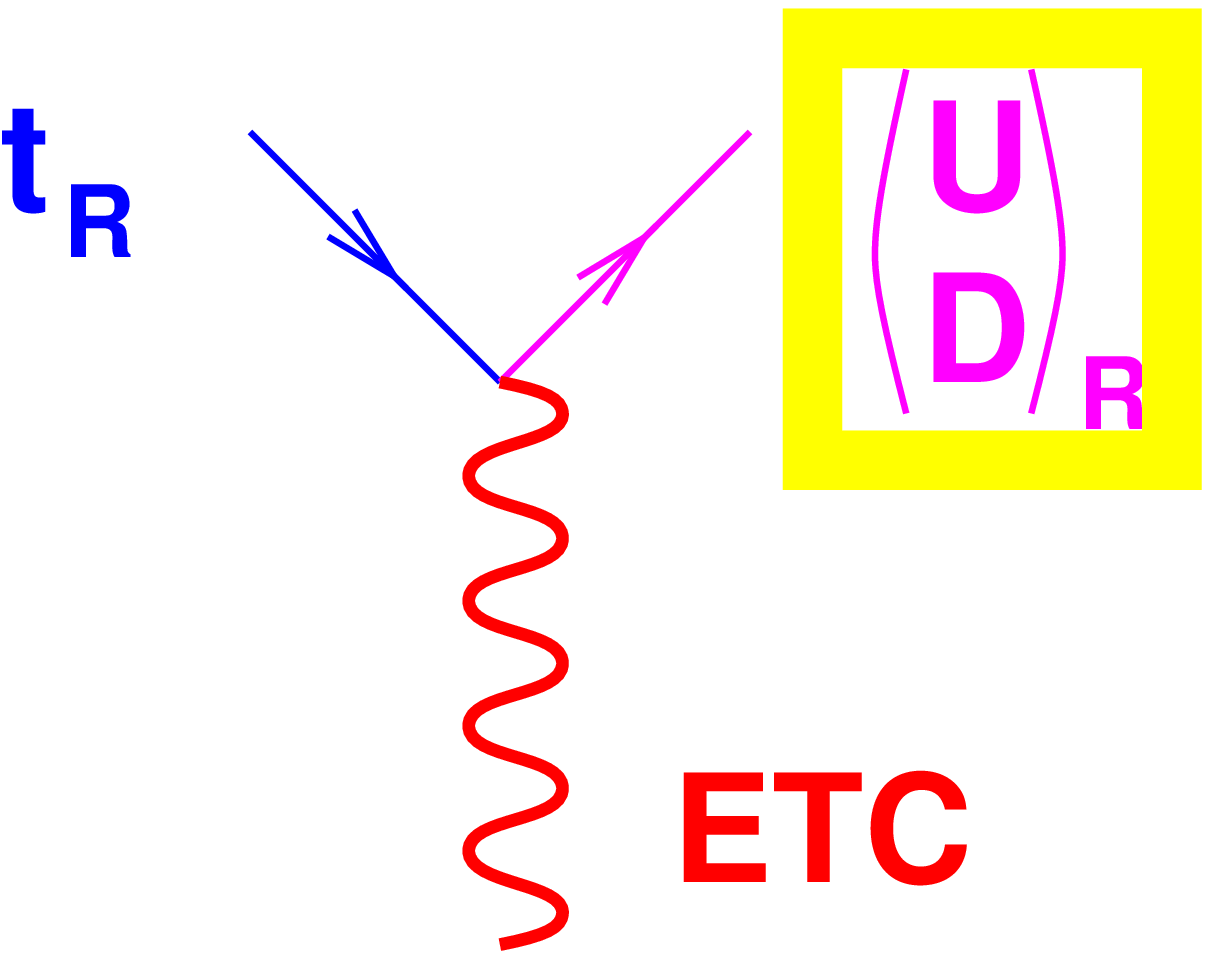}}
\end{center}
\caption{Fermion currents coupling to the weak-doublet ETC boson that
generates $m_t$ in non-commuting ETC models.}
\label{ehs:fig5:ncetc-vv}
\end{figure}

At the same time, there is still a contribution to $R_b$ from the
dynamics that generates $m_t$.  The ETC boson responsible for $m_t$
now couples weak-doublet fermions to weak-singlet technifermions (and
vice versa) as in fig. \ref{ehs:fig5:ncetc-vv}.  The radiative correction
to the $Zb\bar{b}$ vertex is as in fig. \ref{ehs:fig4:etc-vv} except that
the technifermions involved are now $U_L$ with $T_3 = +\frac{1}{2}$.
As a result, the shifts in $\delta g_L$ and $R_b$ have the same size
as the results in eqns. (\ref{ehs:eq:delgg}) and (\ref{ehs:eq:delrbb}) but the
opposite sign.\cite{Chivukula:1994mn}.  Overall, then, the ETC and
$ZZ'$ mixing contributions to $R_b$ in non-commuting ETC models have
equal magnitude and opposite sign, enabling $R_b$ to be consistent
with experiment.  The key element that permits a large $m_t$ and a
small value of $R_b$ to co-exist is the presence of non-standard weak
interactions for the top quark\cite{Chivukula:1994mn}. This is
something experiment can test, and has since been incorporated into
models such as topflavor\cite{Muller:1996dj, Malkawi:1996fs} and top
seesaw.\cite{Burdman:1998vw,He:1999vp}

\begin{figure}[tb] 
\begin{center}
\scalebox{.5}{\includegraphics{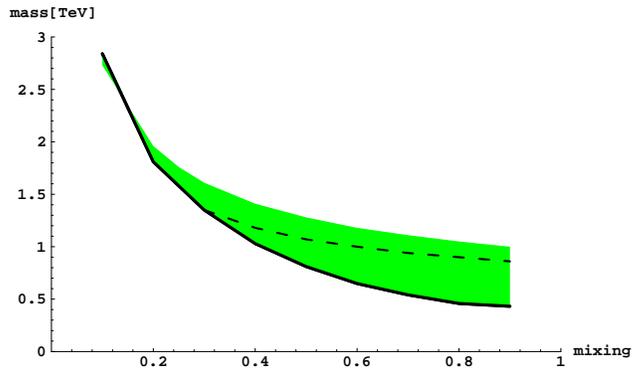}}
\end{center}
\caption{FNAL Run II single top production can explore the shaded
region of the $M_{W'}$ vs. $\sin^2\phi$
plane.\protect\cite{Simmons:1997ws} The area below the solid curve is
excluded by precision electroweak data.\protect\cite{Chivukula:1996gu}
In the shaded region $R_\sigma$ increases by$\geq$ 16\%; below the
dashed curve, by $\geq$ 24\%.}
\label{ehs:fig6:singtp}
\end{figure}

There are several ways to test whether the high-energy weak
interactions have the form $SU(2)_h \times SU(2)_\ell$.  One
possibility is to search for the extra weak bosons.  The bosons'
predicted effects on precision electroweak data gives rise to the
exclusion curve\cite{Chivukula:1996gu} in fig. 
\ref{ehs:fig6:singtp}. Low-energy exchange of $Z^H$ and $W^H$ bosons would
cause apparent four-fermion contact interactions; LEP limits on $eebb$
and $ee\tau\tau$ contact terms imply\cite{Lynch:2000md} $M_{Z^H} \geq
400$ GeV.  Direct production of $Z^H$ and $W^H$ at Fermilab is also
feasible; a Run II search for $Z^H \to \tau\tau \to e\mu X$ will be
sensitive\cite{Lynch:2000md} to $Z^H$ masses up to 650 - 850 GeV.
Another possibility is to measure the top quark's weak interactions in
single top production.  Run II should measure the ratio of single top
and single charged lepton cross-sections $R_\sigma \equiv
\sigma_{tb}/\sigma_{\ell\nu}$ to $\pm 8\%$ in the $W^*$
process.\cite{Heinson:1996pi,Heinson:1997zm,Smith:1996ij} A number of
systematic uncertainties, such as those from parton distribution
functions, cancel in the ratio.  In the Standard Model, $R_\sigma$ is proportional
to the square of the $Wtb$ coupling. Non-commuting ETC models affect
the ratio in two ways: mixing of the $W_h$ and $W_\ell$ alters the
$W^L$ coupling to fermions, and both $W^L$ and $W^H$ exchange
contributes to the cross-sections.  Note that the ETC dynamics which
generates $m_t$ has no effect on the $Wtb$ vertex because the relevant
ETC boson does not couple to $b_R$. Computing the total shift in
$R_\sigma$ reveals (see fig. \ref{ehs:fig6:singtp}) that Run II will be
sensitive \cite{Simmons:1997ws} to $W^H$ bosons up to masses of about
1.5 TeV.

\section{The FCNC Challenge}

In order for extended technicolor to produce the wide range of
observed fermion masses, it is necessary for ETC dynamics to couple
differently to like-charge fermions belonging to different
generations.  This causes ETC boson exchange to contribute to
flavor-changing neutral currents, a potential source of severe
constraints on model-building.  

For example, in the neutral Kaon system, exchange of ETC bosons
contributes to the square of the $K_L K_S$ mass difference by an
amount of order
\begin{equation}
(\Delta M^2_K)_{ETC} \simeq {g^2_{ETC} \, 
{\rm Re}(\theta^2_{sd}) \over {2 M^2_{ETC}}} f_K^2 M_K^2
\end{equation}
where $f_K$ is the Kaon decay constant and $M_K$ is the average
neutral Kaon mass.  Because the experimental upper bound on the Kaon
mass difference is $\Delta M_K < 3.5 \times 10^{-12}$ MeV \cite{ehs_PDBook}, one may
deduce
\begin{equation}
{M_{ETC} \over {g_{ETC} \, \sqrt{{\rm Re}(\theta^2_{sd})}}} > 
  600 {\rm TeV}
\label{ehs:eq:kaonlim}
\end{equation}
Remembering that the fermion mass $m_f$ produced by exchange of an ETC
boson of mass $M_{ETC}$ and coupling $g_{ETC}$ scales as $m_f \sim
g^2_{ETC}/M^2_{ETC}$, we see that the limit (\ref{ehs:eq:kaonlim})
makes it difficult\footnote{An even more stringent limit may be found
in \cite{Lane:2000pa}.} for ETC to produce fermion masses much larger
than an MeV.

In order to address this issue, let us revisit the origin of fermion
masses in ETC models in a little more detail\footnote{This discussion was inspired by refs. \cite{Chivukula:2000mb,Lane:1993wz}.}.  The expression for the
dynamically-generated fermion mass $m_f$ is
\begin{equation}
 m_f \approx {{g_{ETC}^2\over M^2_{ETC}}} \langle\overline{U}
  U\rangle_{ETC}
\end{equation}
where the condensate is evaluated at the ETC scale.  In previous numerical
estimates of the sizes of fermion masses, we have used the
approximation 
\begin{equation}
\langle\overline{U} U\rangle_{ETC} \approx
\langle\overline{U} U\rangle_{TC} \approx 4\pi
F^3_{TC}.
\label{ehs:eq:condapprox}
\end{equation}
More generally, however, the condensate scales as
\begin{equation}
\langle\overline{U} U\rangle_{ETC}
= \langle\overline{U} U\rangle_{TC}
\exp\left(\int_{\Lambda_{TC}}^{M_{ETC}} {d\mu \over \mu}
\gamma_m(\mu)\right)
\label{ehs:eq:integr}
\end{equation}
If the technicolor gauge dynamics resemble those of QCD, the value of
the anomalous dimension $\gamma_m$ is small over the integration
range, so that the integral is negligible and the approximation
(\ref{ehs:eq:condapprox}) holds.

This immediately raises the question: what if the technicolor coupling
instead runs slowly from $\Lambda_{TC}$ up to $M_{ETC}$?  In other
words, what would happen if the technicolor beta function were small,
$\beta_{TC} \approx 0$, making the coupling ``walk'' instead of
running like the QCD coupling?

\begin{figure}[tb]
\begin{center}
\scalebox{.4}{\includegraphics{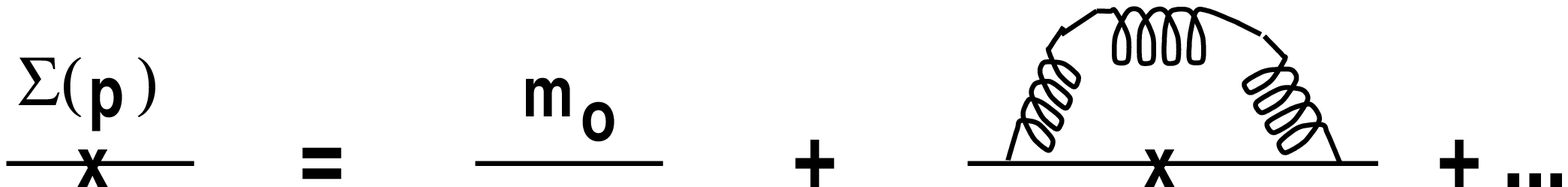}}
\end{center}
\vspace{-.2cm}
\caption{Gap equation for generation of dynamical technifermion mass $\Sigma(p)$.}
\label{ehs:fig7:gap}
\end{figure}

The answer lies in the dynamics by which masses for ordinary fermions
and technifermions arise.  Consider the Schwinger-Dyson gap equation
for the dynamical technifermion mass $\Sigma(p)$ in the rainbow
approximation (fig. \ref{ehs:fig7:gap}).  The phenomenologically
interesting solutions to the gap equation are those manifesting chiral
symmetry breaking: those for which $\Sigma(p) \neq 0$ even if the bare
mass $m_o$ vanishes. Detailed analysis
\cite{Yamawaki:1986zg,Appelquist:1986an,Appelquist:1987fc,Cohen:1989sq,Chivukula:1990bc}
has shown that a chiral symmetry breaking occurs only when the value
of $\alpha_{TC}(\Lambda_{TC})$ approaches the critical value $\alpha_c
\equiv \pi/3 C_2(R)$.  Since the anomalous dimension $\gamma_m(\mu)$
can be written as $1 - \sqrt{1 - {\alpha_{TC}(\mu)}/{\alpha_c}}$, a
large value of $\alpha_TC(\Lambda_{TC})$ implies that
$\gamma_m(\Lambda_{TC}) \approx 1$.  This gives the starting point for
the integration in eqn. (\ref{ehs:eq:integr}).  If $\beta_{TC} \approx 0$,
then $\alpha_{TC} \sim \alpha_c$ and $\gamma_m \sim 1$ persist up to
the scale $M_{ETC}$.  This enhances the integral in
eqn. (\ref{ehs:eq:integr}) and enables ETC to generate fermion masses as
large as $m_s$ or $m_c$:
\begin{equation} 
m_{q,l} = {g^2_{ETC} \over {M^2_{ETC}}} \times
\left(\langle\overline{T}T\rangle_{ETC} \cong 
\langle\overline{T}T\rangle_{TC} \, {{M_{ETC}}
      \over {\Lambda_{TC}}} \right)
\end{equation}
where $M_{ETC}/\Lambda_{TC} \sim 100 - 1000$.

The small technicolor beta function that produces enhanced fermion
masses can arise in models with many technifermions in the standard
vector representation of the technicolor gauge group or in models with
fermions in several different technicolor representations.  In either
case, the chiral symmetry-breaking sector is enlarged relative to that
of minimal technicolor models.  As a result, one expects a
proliferation of technipions and small technipion decay constants
$f_{TC} \ll v$.  At first glance, it appears that the models will
suffer from unacceptably large contributions to S (because of the
large number of technifermions) and from the presence of many light
pseudo-Nambu-Goldstone bosons (PNGBs) which have not been observed.
However, the nature of the strong walking-technicolor dynamics must be
taken into account.  First, QCD is no longer a reliable guide for the
estimation of contributions to S; the walking models have a new
pattern of resonance masses (possibly a tower of $\rho_T$ and
$\omega_T$ states), more flavors, and fermions in non-vectorial gauge
representations \cite{Lane:2000pa}.  In the absence of compelling
estimates of S in walking gauge theories, S does not provide a
decisive test of these models.  Second, the walking dynamics which
enhances the technifermion condensate also enlarges the masses of the
PNGBs
\begin{equation}
F^2_{TC} M^2_{\pi_T} \approx {g^2_{ETC} \over M^2_{ETC}} 
\left(\langle\overline{T}T\rangle_{ETC}\right)^2
\end{equation}
enabling them to meet current experimental constraints.

The phenomenological signatures of walking technicolor have been
studied in models known as ``lowscale
technicolor''\cite{Eichten:1996dx,Eichten:1996dx,Lane:1999uh}.  The
primary signals exploit the contrasting effects of walking on the
masses of the vector mesons $\rho_T$ and $\omega_T$ (which are
reduced) and those of the technipions $\pi_T$ (which are enhanced).
Lighter technivector mesons are more readily produced in colliders,
and if the technipion masses are enhanced enough to close the $\rho_T
\to \pi_T \pi_T$ and $\omega_T \to 3 \pi_T$ decay channels, the
technivectors will quite visibly to final states including electroweak
gauge bosons.  For instance \cite{Lane:1999uh}, if one takes the
number of weak-doublets of technifermions to be $N_D \approx 10$ in
order to induce walking, the technivector meson masses are reduced to
$M_{\rho_T} \approx M_{\omega_T} \approx 2 v / \sqrt{N_D} \approx 200$
GeV.  At the same time, the effects of walking tend to raise the mass
of the $\pi_T$ to over 100 GeV.  The dominant technirho decays will
then be to one $\pi_T$ plus one electroweak gauge boson or to two
electroweak bosons. The $\pi_T$ are expected to decay to $f\bar{f}$
pairs through ETC couplings, making decays to heavy flavors dominate.

Signatures of low-scale technicolor would be visible at both hadron
and lepton colliders.  Current limits have been summarized by
M. Narain in these Proceedings (or see \cite{ehs_PDBook}); these
include searches for $\rho_T \to W \pi_T$ followed by $\pi_T \to
b\bar{b}, c\bar{c} b\bar{c}, c\bar{b}$, for pair-production of
technipions with leptoquark quantum numbers, and for production of
$\rho_T$ and $\omega_T$ decaying to lepton pairs through mixing with
electroweak gauge bosons.  Future experiments will, naturally, have
greater reach.  For example, a 200 GeV muon collider could resolve
\cite{Eichten:1998kn} the peaks of even nearly degenerate $\rho_T$ and
$\omega_T$ in the process $\mu^+ \mu^- \to \rho_T, \omega_T \to e^+
e^-$.  As another example, the LHC, technirhos with masses up to 500
or more GeV would provide a visible signal in 30 fb$^{-1}$ of ATLAS
data through decays to a $WZ$ pair which then decay to leptons
\cite{Witzeling:1999yw}.  Summaries of the expected reach of various
 technicolor searches at Tevatron Run II and the LHC may be
found in refs. \cite{ehs_PDBook} and \cite{Chivukula:1995dt},
respectively.

\section{The $\Delta\rho$ Challenge}

At tree-level in the Standard Model, $\rho \equiv M_W^2 / M_Z^2 \cos^2\theta_W
\equiv 1$ due to a ``custodial'' global $SU(2)$ symmetry relating
members of a weak isodoublet.  Because the two fermions in each
isodoublet have different masses and hypercharges, however, oblique
radiative corrections to the $W$ and $Z$ propagators alter the value
of $\rho$.  The one-loop correction from the (t,b) doublet is
particularly large because $m_t \gg m_b$.  Experiment\cite{ehs_PDBook}
finds $\vert\Delta\rho\vert \leq 0.4\% $, a stringent constraint on
isospin-violating new physics.

\begin{figure}[tb]
\begin{center}
\scalebox{.4}{\includegraphics{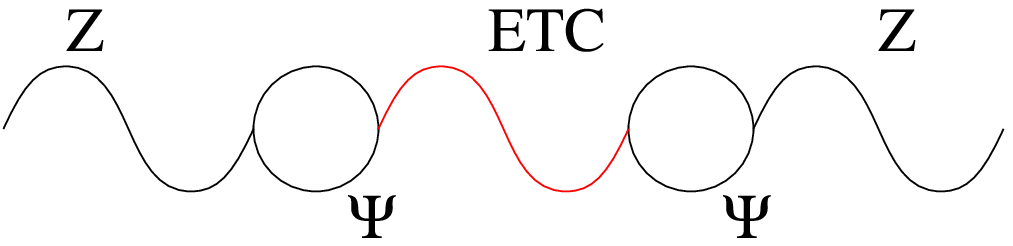}}
\hspace{1cm} \scalebox{.4}{\includegraphics{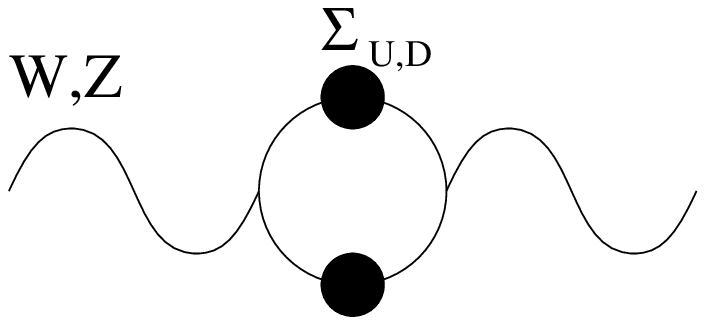}}
\end{center}
\vspace{-.2cm}
\caption{ ETC contributions to $\Delta\rho$: (left) direct, from gauge
boson mixing (right) indirect, from technifermion mass splitting.}
\label{ehs:fig8:zetcmix}
\end{figure}

\begin{figure}[tb]
\begin{center}
\scalebox{.45}{\includegraphics{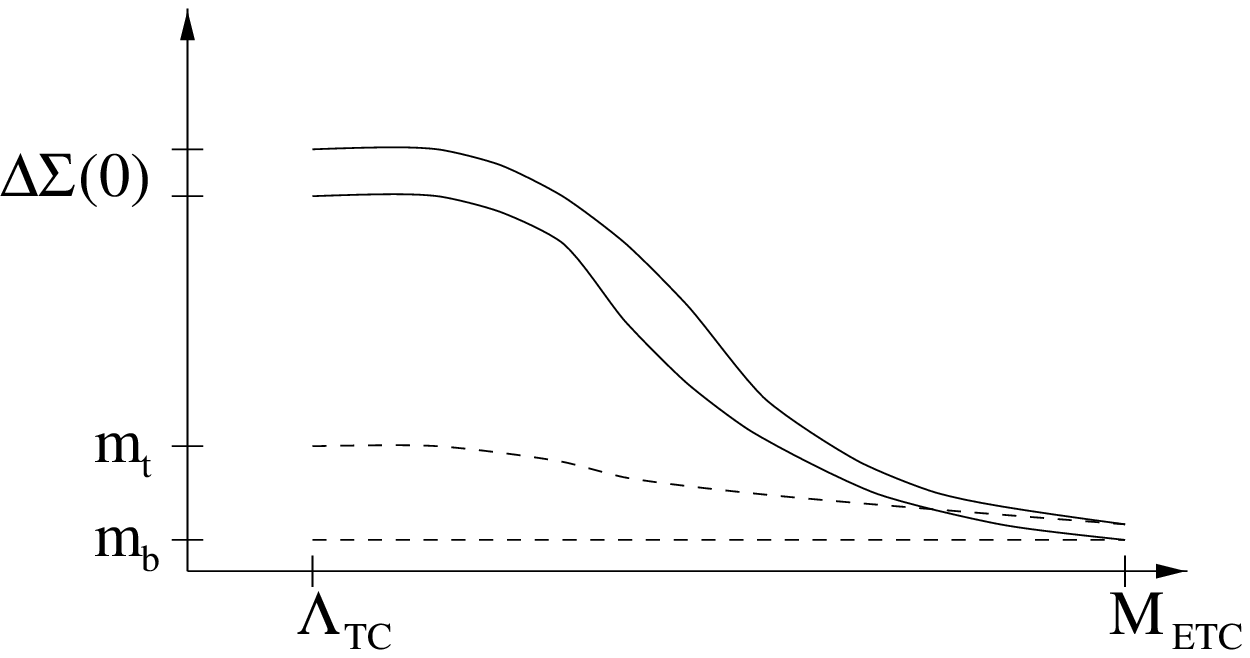}} 
\scalebox{.45}{\includegraphics{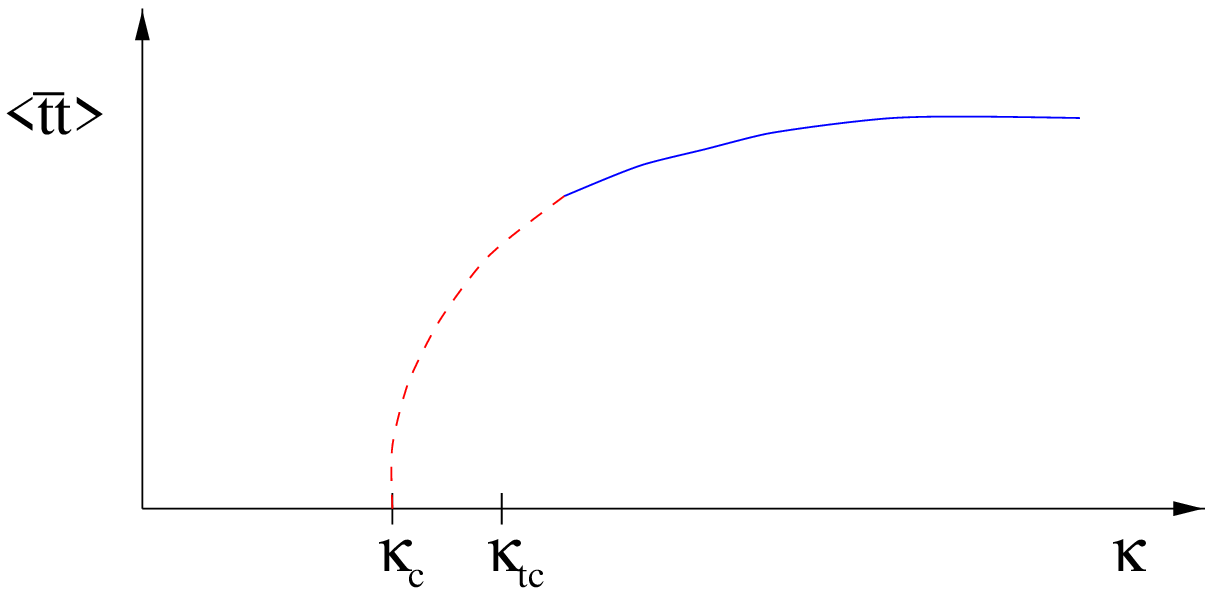}}
\end{center}
\caption{(left) ETC and new top dynamics generate masses for
technifermions, $t$ and $b$. (right) Second-order phase transition forms a
top condensate for $\kappa > \kappa_c$.}
\label{ehs:fig9:dynmasss}
\end{figure}

Dynamical theories of mass generation like ETC must break weak isospin
in order to produce the large top-bottom mass splitting.  However, the
new dynamics may also cause additional, large contributions to
$\delta\rho$.  Direct
mixing between and ETC gauge boson and the Z (fig. \ref{ehs:fig8:zetcmix})
induces the dangerous effect\cite{Appelquist:1984nc,Appelquist:1985rr}
\begin{equation}
\Delta\rho \approx  12\% \cdot 
\left({\sqrt{N_D} F_{TC} \over 250 {\rm\ GeV}}\right)^2
\cdot \left({1 {\rm\ TeV} \over M_{ETC}/g_{ETC}}\right)^2
\end{equation}
in models with $N_D$ technifermion doublets and technipion decay
constant $F_{TC}$.  To avoid this, one could make the ETC boson heavy;
however the required $M_{ETC}/g_{ETC} >\ 5.5{\rm TeV} (\sqrt{N_D}
F_{TC} / 250\ {\rm GeV})$ is too large to produce $m_t = 175$ GeV.
Instead, one must obtain $N_D F^2_{TC} \ll (250{\rm GeV})^2$ by
separating the ETC sectors responsible for electroweak symmetry
breaking and the top mass.  A second contribution comes
indirectly\cite{Chivukula:1988qr} through the technifermion mass splitting:
$\Delta\rho \sim (\Sigma_U(0) - \Sigma_D(0))^2/M_Z^2$, as in fig. 
\ref{ehs:fig8:zetcmix}.  Again, a cure\cite{Hill:1991at,Martin:1992uz,Martin:1992aq,Martin:1993mj,Lindner:1992bs,Bonisch:1991vd,Hill:1993ev,Hill:1995hp,Chivukula:1998vd,Chivukula:1998uf,Chivukula:1990bc} is to arrange
for the $t$ and $b$ to get only part of their mass from technicolor.
As sketched in fig. \ref{ehs:fig9:dynmasss}, suppose $M_{ETC}$ is large
and ETC makes only a small contribution to the fermion and
technifermion masses.  At a scale between $M_{ETC}$ and $\Lambda_{TC}$
new strong dynamics felt only by (t,b) turns on and generates $m_t \gg
m_b$.  The technifermion mass splitting is small, $\Delta\Sigma(0)
\approx m_t(M_{ETC} - m_b(M_{ETC}) \ll m_t$, and no large
contributions to $\Delta\rho$ ensue.

The realization that new strongly-coupled dynamics for the (t,b)
doublet could be so useful has had a dramatic effect on
model-building.  Models in which some (topcolor\cite{Hill:1991at,Martin:1992uz,Martin:1992aq,Martin:1993mj,Lindner:1992bs,Bonisch:1991vd,Hill:1993ev,Hill:1995hp,Chivukula:1998vd,Chivukula:1998uf}) or even all
(top mode \cite{Miransky:1989xi,Miransky:1989ds,Nambu:1989jt,Marciano:1989xd,Bardeen:1990ds}, top seesaw\cite{Dobrescu:1998nm,Chivukula:1998wd}) of electroweak symmetry breaking is due to a top condensate have proliferated.  One
physical realization of a new interaction for the top is a
spontaneously broken extended gauge group called topcolor\cite{Hill:1991at,Martin:1992uz,Martin:1992aq,Martin:1993mj,Lindner:1992bs,Bonisch:1991vd,Hill:1993ev,Hill:1995hp,Chivukula:1998vd,Chivukula:1998uf}:
$SU(3)_h \times SU(3)_\ell \to SU(3)_{QCD}$.  The (t,b) doublet
transforms under $SU(3)_h$ and the light quarks, under $SU(3)_\ell$.
Below the symmetry-breaking scale $M$, the spectrum includes massive
topgluons which mediate vectorial color-octet interactions among top
quarks: $ -(4\pi\kappa/M^2) (\bar{t}\gamma_\mu \frac{\lambda^a}{2}
t)^2$.  If the coupling $\kappa$ lies above a critical value
($\kappa_c = 3\pi/8$ in the NJL\cite{Nambu:1961tp,Nambu:1961fr}
approximation), a top condensate forms (fig. \ref{ehs:fig9:dynmasss}).
For a second-order phase transition, $\langle \bar{t} t \rangle / M^3
\propto (\kappa - \kappa_c) / \kappa_c$, so the top quark mass
generated by this dynamics can lie well below the symmetry breaking
scale; so long as $M$ is not too large, the scale separation need not
imply an unacceptable degree of fine tuning.

A more complete model incorporating these ideas is topcolor-assisted
technicolor\cite{Hill:1991at,Martin:1992uz,Martin:1992aq,Martin:1993mj,Lindner:1992bs,Bonisch:1991vd,Hill:1993ev,Hill:1995hp,Chivukula:1998vd,Chivukula:1998uf} (TC2).  The symmetry-breaking structure is:
\begin{eqnarray} 
G_{TC} &\times& {SU(3)_h \times SU(3)_\ell} \times SU(2)_W
  \times  {U(1)_h \times U(1)_\ell} \nonumber \\
&\downarrow&\ \ \ M \geq 1\ {\rm TeV} \nonumber \\
G_{TC} &\times& {SU(3)_{QCD}}  \times SU(2)_{W} \times
  { U(1)_Y} \nonumber \\
&\downarrow&\ \ \ \Lambda_{TC}\sim 1\ {\rm  TeV}\nonumber \\
G_{TC} &\times& SU(3)_{QCD} \times {U(1)_{EM}}
\end{eqnarray}
Below the scale $M$, the heavy topgluons and Z' mediate
new effective interactions\cite{Hill:1991at,Martin:1992uz,Martin:1992aq,Martin:1993mj,Lindner:1992bs,Bonisch:1991vd,Hill:1993ev,Hill:1995hp,Chivukula:1998vd,Chivukula:1998uf,Chivukula:1995dc,Lane:1995gw,Lane:1996ua,Buchalla:1996dp,Lane:1998qi,Popovic:1998vb} for the (t,b) doublet
\begin{equation} 
-{4\pi \kappa_3\over{M^2}}\left[\overline{\psi}\gamma_\mu  
{{\lambda^a}\over{2}} \psi \right]^2 - {4\pi
\kappa_1\over{M^2}}
\left[{1\over3}\overline{\psi_L}\gamma_\mu  \psi_L
+{4\over3}\overline{t_R}\gamma_\mu  t_R -{2\over3}\overline{b_R}\gamma_\mu
b_R \right]^2 
\label{ehs:eq:tc2outline}
\end{equation}
where the $\lambda^a$ are color matrices and $g_{3h} \gg g_{3\ell}$,
$g_{1h} \gg g_{1\ell}$.  The $\kappa_3$ terms are uniformly attractive;
were they alone, they would generate large $m_t$ {\bf and} $m_b$.  The
$\kappa_1$ terms, in contrast, include a repulsive component for $b$.  As a
result, the combined effective interactions\cite{Hill:1991at,Martin:1992uz,Martin:1992aq,Martin:1993mj,Lindner:1992bs,Bonisch:1991vd,Hill:1993ev,Hill:1995hp,Chivukula:1998vd,Chivukula:1998uf,Chivukula:1995dc,Lane:1995gw,Lane:1996ua,Buchalla:1996dp,Lane:1998qi,Popovic:1998vb}
\begin{equation}
\kappa^t = \kappa_3 +{1\over3}\kappa_1 >
\kappa_c  > \kappa_3 -{1\over 6}\kappa_1 =\kappa^b
\end{equation}
can be super-critical for top, causing $\langle\bar{t}t\rangle \neq 0$ and
a large $m_t$, and sub-critical for bottom, leaving $\langle\bar{b}b\rangle
= 0$. 

The benefits of including new strong dynamics for the top quark are
clear in TC2
models.\cite{Chivukula:1995dc,Lane:1995gw,Lane:1996ua,Buchalla:1996dp,Lane:1998qi,Popovic:1998vb} 
Because technicolor is responsible for most of electroweak symmetry
breaking, $\Delta\rho \approx 0$.  Direct contributions to
$\Delta\rho$ are avoided because the top condensate provides only
$f\sim 60$ GeV; indirect contributions are not an issue if the
technifermion hypercharges preserve weak isospin.  The top condensate
yields a large top mass.  ETC dynamics at $M_{ETC} \gg 1$ TeV generate
the light $m_f$ without large FCNC and contribute only $\sim 1$ GeV to
the heavy quark masses so there is no large shift in $R_b$.

Three classes of models of new strong top dynamics with distinctive
spectra are known as topcolor\,\cite{Hill:1991at,Hill:1995hp},
flavor-universal extended
color\,\cite{Chivukula:1996yr,Popovic:1998vb, Lane:1998qi}, and top
seesaw\,\cite{Dobrescu:1998nm}. Exotic particles in these models
include colored gauge bosons (topgluons, colorons), color-singlet
gauge bosons (Z'), composite scalar states (top-pions, q-pions), and
heavy fermions (usually, but not
always\,\cite{Burdman:1998vw,He:1999vp}, weak singlets).  Because
strong top dynamics is the subject of a talk by B. Dobrescu in these
proceedings, it suffices here to note briefly that numerous searches
for these new states have been attempted or proposed. For example, CDF
has searched \cite{Abe:1998uz} for topgluons and Z' in heavy quark
final states, and the potential for finding $Z' \to \tau\tau \to e
\mu$ at Run II has been discussed in \cite{Lynch:2000md}. Limits on
flavor-universal colorons are summarized in
ref. \cite{Bertram:1998wf}. The phenomenology of the new weak-singlet
quarks present in top seesaw models has been discussed in
refs. \cite{Collins:1999rz,Popovic:2000dx}.  Searches for the
composite scalars of strong top dynamics models are analogous to the
widely discussed methods for finding the extra scalars in
multiple-Higgs models.

\section{Summary}

\begin{figure}[tb]
\begin{center}
\rotatebox{270}{\scalebox{.17}{\includegraphics{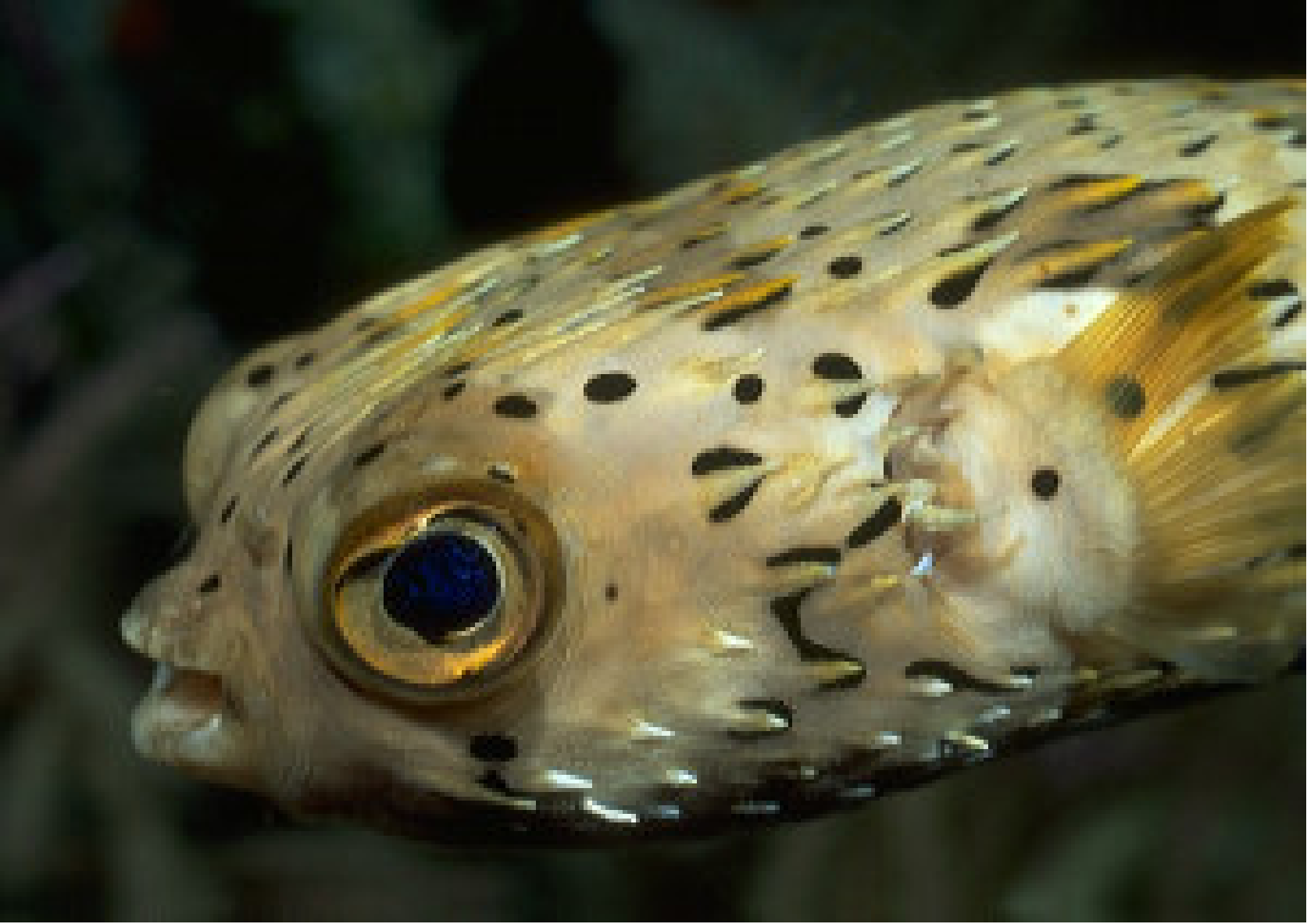}}}
\hspace{.2cm}\rotatebox{270}{\scalebox{.177}{\includegraphics{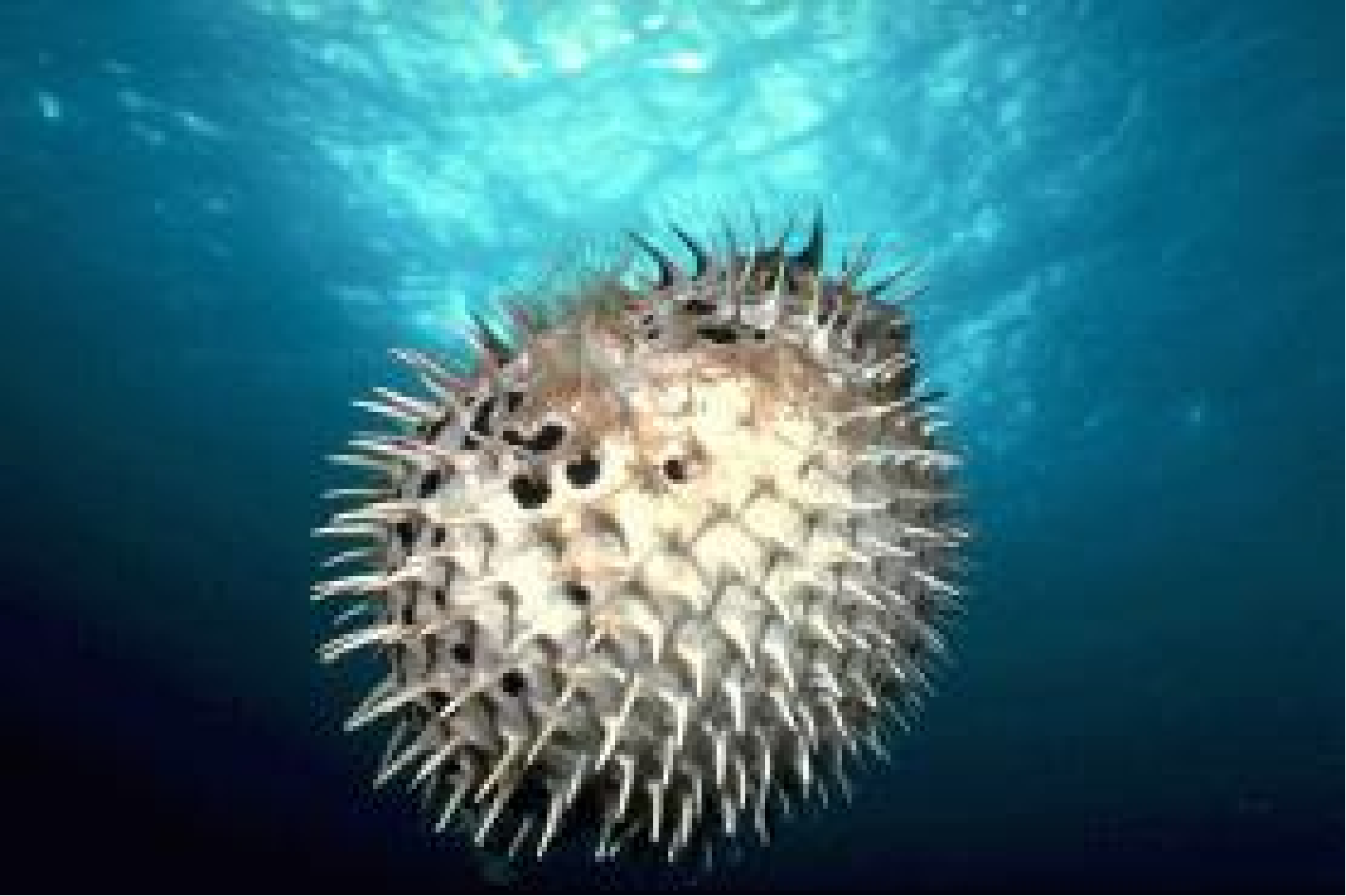}}}
\hspace{.2cm} \lower99pt\hbox{\scalebox{.32}{\includegraphics{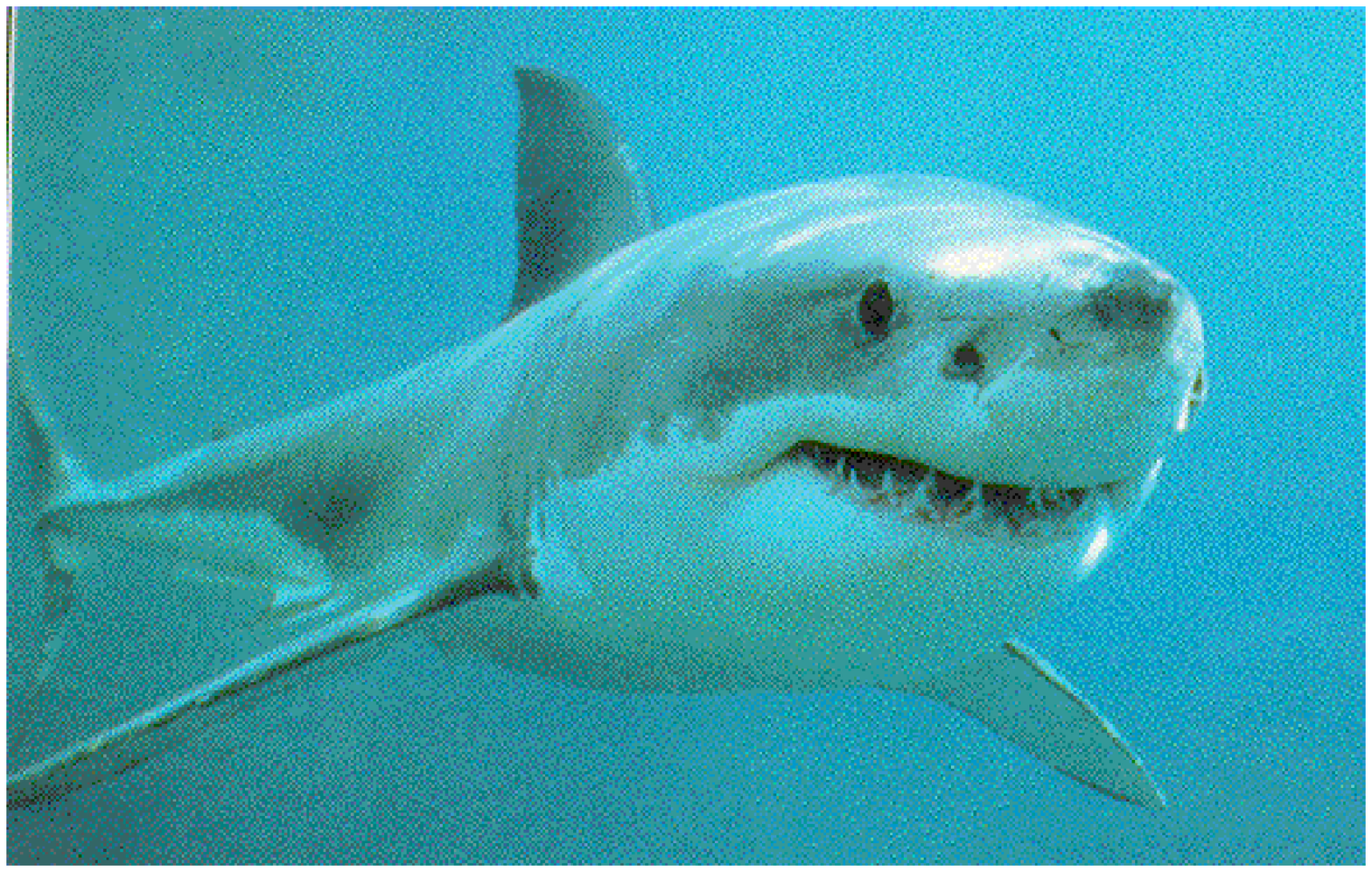}}}
\end{center}
\caption{(a) Puffer fish. (b) Inflated spiny puffer fish. (c) Great
white shark.}
\label{ehs:fig10:fishies}
\end{figure}

Dynamical symmetry breaking models, such as extended technicolor, use
familiar gauge dynamics to give form to that most elusive quarry
(fig. \ref{ehs:fig10:fishies}a), the origin of mass.  Modern dynamical
theories such as non-commuting ETC, low-scale technicolor, or
topcolor-assisted technicolor have evolved in response to the
pressures applied by increasingly precise measurements of observables
such as $R_b$, flavor-changing neutral currents, and $\Delta\rho$.
All of these theories offer intriguing and distinctive signatures
(fig. \ref{ehs:fig10:fishies}b), many of which are discussed in these
Proceedings by M. Narain.  As the next round of collider experiments
(fig. \ref{ehs:fig10:fishies}c) begins, I wish them ``Good Hunting!''

% figures should be put into the text as floats.
% Use the graphicx package (distributed with LaTeX2e).
% See the LaTeX Graphics Companion by Michel Goosens, Sebastian Rahtz,
% and Frank Mittelbach for instance.
%
% Here is an example of the general form of a figure:
% Fill in the caption in the braces of the \caption{} command. Put the label
% that you will use with \ref{} command in the braces of the \label{} command.
%
% \begin{figure}
% \includegraphics{}%
% \caption{}
% \label{}
% \end{figure}

% Create the reference section using BibTeX:
\bibliography{fish}

\end{document}